\newcommand{\ednote}[1]{\marginpar{\texttt{\tiny #1}}}
\renewcommand{\ednote}[1]{}
\newcommand{\completion}[2]{\mathcal{C}_{#1}^{#2}}

\newcommand{\adjoin}{\rightharpoonup}
\newcommand{\card}[1]{\left| #1 \right|}

\newcommand{\mcup}{+}
\newcommand{\msub}{\leq}
\newcommand{\pay}{p}
\newcommand{\quotpay}{\pay'}

\newcommand{\wwwMml}{http://mizar.uwb.edu.pl/version/current/mml/}
\newcommand{\wwwMizar}{http://www.mizar.org}
\newcommand{\wwwGithub}{https://github.com/formare/auctions/}

\newcommand{\wwwMbc}{http://caminati.tk/}
\newcommand{\myTitle}{Budget Imbalance Criteria for Auctions: A Formalized Theorem}

\newcommand{\he}{she}

\newcommand{\his}{her}

\newcommand{\himself}{herself}

\newcommand{\I}{Isabelle}
\newcommand{\fo}{ForMaRE}
\newcommand{\pred}{adequate}
\newcommand{\query}[1]{\marginnote{\raggedright\footnotesize\itshape\hrule\smallskip{#1}\smallskip\hrule}}
\renewcommand{\query}[1]{} 
\newcommand{\rnote}[1]{\query{#1}}
\newcommand{\lnote}[1]{\reversemarginpar\query{#1}\normalmarginpar}
\newcommand{\vv}[1]{\overline{#1}}
\newcommand{\out}{-}
\newcommand{\sdiff}{\backslash}
\newcommand{\Pay}{P}
\newcommand{\Set}[1]{\left\{ \left| #1 \right| \right\}}
\newcommand{\RRange}[1]{\left\| #1 \right\|}
\newcommand{\ul}{\underline}
\newcommand{\ol}{\overline}
\newcommand{\cex}{counterexample}
\newcommand{\nb}{NB}
\renewcommand{\v}[1]{\vec{#1}}

\documentclass[
oneside
]{llncs}

\usepackage{graphicx}
\usepackage{alltt}
\usepackage{amsmath} 
\usepackage{amssymb}

\usepackage[natbib=true
,backend=bibtex,maxbibnames=100,maxcitenames=3]{biblatex} 
\usepackage[symbol]{footmisc}
\usepackage[]{marginnote}
\usepackage[]{url}
\DeclareMathOperator{\dom}{dom}
\DeclareMathOperator{\ran}{rng}

\urldef{\WwwMbc}\url{\wwwMbc}
\urldef{\WwwMizar}\url{\wwwMizar}
\urldef{\WwwMml}\url{\wwwMml}
\urldef{\WwwGithub}\url{\wwwGithub}

\author{
}

\begin{document}


\title{\myTitle{}\thanks{This work has been supported by EPSRC grant EP/J007498/1 and an LMS Computer Science Small Grant.}}

\author{
Marco B.\ Caminati\inst{1}
\and Manfred Kerber\inst{1}
\and Colin Rowat\inst{2}
}

\institute{%
School~of~Computer~Science, University of Birmingham, UK%
\and Economics, University of Birmingham, UK\\
Project homepage: \url{http://cs.bham.ac.uk/research/projects/formare/}
}

\maketitle

\begin{abstract}
We present an original theorem in auction theory: it specifies general conditions under which the sum of the payments of all bidders is necessarily \ednote{deleted: granted, added: necessarily} not \ednote{deleted: to be} identically zero, and more generally not constant.
Moreover, it explicitly supplies a construction for a finite minimal set of possible bids on which such a\ednote{added `a'} sum is not constant.
In particular, this theorem applies to the important case of a second-price Vickrey auction, where it reduces to a basic result \ednote{changed results to result} of which a novel \ednote{deleted `, alternative'} proof is \ednote{deleted: thus} given.
To enhance the confidence in this new theorem, it has been formalized in \I{}/HOL: 
the main results and definitions of the formal proof are reproduced here in common mathematical language, and are\ednote{added `are'} accompanied by \ednote{repalaced `with' by `by'} an informal discussion about the underlying ideas.
\end{abstract}


\section{Introduction} \label{sec:introduction}
\label{RefIntro}
The \fo{} project \cite{LRK:FormareProject13:short} employs formal methods to provide a\ednote{changed `an' to `a'} unified
\lnote{MC: `unitary' -> `unified' as requested by R2}
approach to both the generation of \ednote{changed: generate} executable code for running auctions and the proving of \ednote{changed: prove} theorems about them.
In this paper, we will describe the formalization of a classical result about the second price Vickrey auction (which will be introduced in Section \ref{RefSectContext}), stating that the sum of the payments for all participants cannot be zero for all possible bids. 
We will indicate this result as \nb{} (for \emph{n}o~\emph{b}alance).

The proof mechanism we present for \nb{} is, to the best of our knowledge, \ednote{deleted `completely'} new.
Although it is also applicable to the specific case of the Vickrey auction, our proof works for a wide class of possible auction mechanisms: indeed, it gives a characterization of the kinds of auctions for which it holds.
By contrast, all the existing proofs we know of vitally rely on the particular algebraic form that the mechanism assumes in the particular case of the Vickrey auction.
Furthermore, our proof explicitly constructs a minimal, finite set of possible bids on which the sum of all the payments is not a constant function.

All the results have been formalized and checked in the \I{}/HOL theorem prover \cite{Nipkow-Paulson-Wenzel:2002}.
\ednote{deleted `However, also'} Because the results are new, they are stated here in common mathematical language, which should be friendlier for the reader.
\ednote{Reformulated sentence}
The lemmas, definitions, and theorems in this paper correspond as far as possible their formalized counterparts and for each statement we indicate the corresponding \I{} name in typewriter font.
The relevant \I{} theory file is \verb|Maskin3.thy|, available at \url{https://github.com/formare/auctions/}.

\subsection{Structure of the paper}

In Section \ref{RefSectContext}, some context is given: we will see the basic mathematical definitions for auctions, which will be needed to state the main theorem \nb{}, the object of the present formalization.

Section \ref{RefSectIdea} informally illustrates the idea behind our original proof of \nb{}, and Section \ref{RefSectLemmaInductive} presents the formal result implementing this idea, which is Lemma \ref{RefLemmaInductive}.
This lemma is the cornerstone of the whole formalization effort presented here: all the other results depend on it.

This fact is illustrated in Section \ref{RefSectVickrey}, where it is informally explained how Lemma\ednote{Capitalized Lemma, but did not do throughout the paper. We should discuss whether we want that.} \ref{RefLemmaInductive} can be applied to the particular case of the Vickrey auction.

This informal explanation is then made formal and systematic in Section \ref{RefSectTheorem}, where ancillary lemmas and definitions are given in order to formally derive from Lemma \ref{RefLemmaInductive} the \ednote{replaced wanted} main result, Theorem \ref{RefThmNbFormal}, which is the formal statement of our generalized version of \nb{}.

\subsection{Notations}
\label{RefSectNotation}
\begin{itemize}
\item
We will represent any function (e.g., the one associating to each participant \his{} bid) in a set-theoretical fashion;
that is, as the set $ \left\{ \left( x, f \left( x \right) \right)| x \in \dom f \right\}$, commonly called the graph of $f$.
Hence, for example, the cartesian product $X \times \left\{ y \right\}$ is the constant function defined on $X$ and yielding $y$.
\item
Similarly, any relation will be represented as the set of the pairs of elements related through it; formally, this means that any relation is a subset of some cartesian product $X \times Y$.
\item
Given a relation $R$, $R \left( X \right)$ will be the image of the set $X$ through $R$: $ R \left( X \right) := 
\left\{ y | \left( x, y \right) \in R \text{ and } x \in X \right\}$. 
For example, given a function $f$, $f^{-1} \left( \left\{ y \right\} \right) $ will then be the fiber of the point $y$ under $f$. 
\item
The restriction of a function $f$ to the set-theoretical difference $\dom f \sdiff X$ will be written $f \out X$;
moreover, in the special case of $X = \left\{ x \right\}$, we will often abuse the notation, writing $f \out x$ instead of $f \out \left\{ x \right\}$.
\item
A multiset (also called a bag) will be extensionally denoted writing, e.g., $ \Set {0,0,1,1,1,2}$: we recall that a multiset is similar to a set, but each member has a multiplicity describing how many times it appears in the multiset.
We will use $\mcup$ as the infix symbol for pairwise multiset union; we will write $A \msub B$ to express the fact that $A$ is a sub-multiset of $B$: for instance, $ \Set{2,1,1} \msub \Set{0,0,1,1,1,2}$ is true. 
\item
Finally, $x \adjoin X$ denotes the operation of union of $x$ with each set in the family of sets $X$:
\begin{align*}
x \adjoin X := \left\{ x \cup x': x' \in X \right\}.
\end{align*}
We will need this operation to state Lemma \ref{RefLemmaInductive}.
\end{itemize}

\section{Statement of the main result}
\label{RefSectContext}
An auction mechanism is mathematically represented through a pair of functions $ a, p $: the first describes how the goods at stake are allocated among the bidders (also called participants or agents), while the second specifies how much each bidder pays following this allocation.
Each possible output of this pair of functions is referred to as an \emph{outcome} of the auction.
Both functions take the same argument, which  
is another function, commonly called a \emph{bid vector}; it describes how much each bidder prefers each possible outcome of the auction. 
This preference relation is usually expressed through money, hence the bid vector associates some outcome to how much the participant values that outcome.
To stick with traditional notation, we will use bold face for bid vectors, as in $ a \left( \v b \right)$.

In the case of a single good being auctioned, the bid vector simply associates to each bidder the amount \he{} bids for the good.
Given a fixed bidder $i$, moreover, $a_i$ is $\left\{ 0,1 \right\}$-valued, corresponding to the fact that $i$ either wins the item or not.
For a single good auction, the Vickrey mechanism has a special relevance because of the formal properties it enjoys \cite{mas-04, la-ca-ke-mo-ro-we-ww-13}.
It works by assigning the good to the highest bidder. 
Each agent then pays a `fee{}' term $\quotpay_i \left( \v b \out i \right)$ irrespective of the outcome; this fee does not depend on how much $i$ \himself{} bids, but only on the other participants' bids: 
hence the argument of $\quotpay_i$ is $\v b \out i$ rather than the full $\v b$.
Additionally, the winner pays a proper price term given by the highest of the bids computed on the set of participants excluding the winner \himself{} (second price); given a fixed bid vector $\v b$, we will denote this amount as $f_2 \left( \v b \right)$.

An often desirable property of an auction mechanism is that it achieves \emph{budget balancing} \cite[Section~2.3]{mil-04}.
This means that the sum of the money given or received by each participant \emph{always} totals to zero:
\begin{align}
\label{RefBalanceDef}
\forall \v b && \sum_{i} \pay_i \left( \v b \right) = 0.
\end{align}
Such a property becomes attractive when 
\begin{quote}
it is preferable to maintain as much wealth as possible within the group of agents, and the transfer payment can be viewed as an undesirable ``cost of implementation''.
\cite
{cavallo2006optimal}
\end{quote}

\noindent\ednote{Added `noindent'}There are important cases in which \eqref{RefBalanceDef} assumes a more specific form:%
\footnote{We recall that bid vectors are modeled as functions, hence we can write $ \v b \out i$, using the notation introduced in Section~\ref{RefSectNotation}.}
\begin{align}
\label{RefBalanceDefParticular}
\forall \v b  && \vv f \left( \v b \right) + \sum_{i} \quotpay_i \left( \v b \out i \right) = 0,
\end{align}
where $\vv f$ typically extracts some kind of maximum: e.g., for the single-good Vickrey
mechanism, $\vv f \left( \v b \right)$ is the second price $f_2 \left( \v b \right)$.
The function $\quotpay_i$ is related to $\pay_i$ through a simple construction we are not interested in\ednote{deleted `,'} here.
Here, the important fact is the formal difference between $\pay_i$ and $\quotpay_i$: the former takes as an argument the full bid vector, while the latter takes a \emph{reduced} bid vector, in which the bid pertaining participant $i$ has been removed.

A standard result \cite[Theorem~2.2]{mil-04}, \cite[Proposition~3]{mas-04}, \cite[Theorem~4.1]{la-ma-80} 
\rnote{I think you should give more intuition. Maybe you do in the Intro. Why would we want budget balancing? Why is this important, relevant ...? 
MC: For the first submission I gave this point up. 
If there is time for a second shot, I will add this.} 
is that for such cases, budget balancing cannot be satisfied: \eqref{RefBalanceDef} is false.
Its known proofs, however, all exploit the particular form of the map $\vv f$ appearing in \eqref{RefBalanceDefParticular} in the considered case, namely that of $\vv f$ being $f_2$.
Vice versa, we will see a proof starting from \eqref{RefBalanceDefParticular} where the latter map is considered as an unknown; the proof will work out the conditions it needs to impose on that map: only after that we will ascertain they are fulfilled for the given cases we are interested in (e.g., the mentioned single-good Vickrey auction).
To be even more informative, the proof will 
show that to falsify equation \eqref{RefBalanceDefParticular}, it is not needed to quantify it over all the possible $\v b$ admissible as an argument to $\vv f$: a smaller, finite set $X$ will be suggested by the proof itself.

Hence, we will consider the logical formula
\begin{align}
\label{RefAlgebraicProblem}
\forall \v b \in X && \sum_{i} \quotpay_i \left( \v b \out i \right) = f \left( \v b \right) 
\end{align}
and study for what $X$ and $f$ it leads to a contradiction (which will include, of course, our starting case \eqref{RefBalanceDefParticular}, where $f = - \vv f$).
%
Since we are going to set up a proof mechanism minimizing the requirements on the generic $f$ (e.g., we are not going to expect that $f$ is the maximum or the second maximum of the bids), 
we must impose some different premises to carry through \ednote{replaced `on' by `through'} a proof.
The main assumption needed 
is one of symmetry: while the $\pay_i$s in \eqref{RefBalanceDef} (and hence the $\quotpay_i$s\ednote{deleted apostrophe twice} in \eqref{RefBalanceDefParticular}) are completely arbitrary, we \ednote{deleted `will'} need to require that they do not depend on re-labeling the participants:
\begin{align}
\label{RefSymmetry}
\exists \Pay && \forall i \  \v b && \quotpay_i \left( \v b \right) = \Pay (\RRange{\v b}),
\end{align}
where $\RRange{R}$
is the \emph{multiset} (or \emph{bag}) obtained from the relation $R$: that is, the range of $R$, but taking into account the multiplicities of possibly repeated values in it.%
\footnote{For example, given the map associating to participant $1$ the bid $10$, to participant $2$ the bid $20$, and to participant $3$ the bid $10$, the obtained multiset is $\Set{10,10,20}$.}
This means that the price paid by any participant $i$ is given by a rule depending only on the amount of each bid other than $i$'s, and not on \emph{who} placed those.
Moreover, such a\ednote{added `a'} rule itself must not depend on $i$.

With this assumption in place, \eqref{RefAlgebraicProblem} becomes
\begin{align}
\label{RefAlgebraicProblSymmetric}
\forall \v b \in X && 
\sum_{i} \Pay \left( \RRange{\v b \out i} \right) = f \left( \v b \right).
\end{align}

\section{Proof idea}
\label{RefSectIdea}
Let $N$ be the number of bidders. 
\rnote{Is this now about proving the generalized theorem with a new proof idea, extending the standard approach? If yes, say so explicitly. Also make it clear of whether the proof is a generalisation of the standard result, or whether the proof idea is new as well. 
MC: Addressed this in Intro and Conclusions.}
The proof starts by considering the case that they all submit the same (fixed but arbitrary) bid $b_0$, whence \eqref{RefAlgebraicProblSymmetric} yields:
\begin{align}
\label{RefIteration0}
\Pay \left( \Set{
\underbrace{b_0, \ldots, b_0}_{N-1}
} \right) 
= k_0 f \left( 
\underbrace{b_0, \ldots, b_0}_{N} 
\right),
\end{align}
with $k_0 := \frac{1}{N}$ not depending on $b_0$.
Then we continue with a $\v b$ in which only one component (let us say the first, for example) assumes an arbitrary value $b_1$ different than $b_0$; thus, \eqref{RefAlgebraicProblSymmetric} gives
\begin{align}
\label{RefIteration1}
(N-1) P \left( \Set{
b_1, \underbrace{b_0, \ldots, b_0}_{N-2}
} \right) = -P \left(\Set{  
\underbrace{b_0, \ldots, b_0}_{N-1}
}\right) - f \left( b_1, b_0, \ldots, b_0 \right).
\end{align}

At this point, we would like to trigger an iterative mechanism by exploiting \eqref{RefIteration0} inside \eqref{RefIteration1}.
A natural imposition to make this possible is to ask that

\begin{align}
\label{RefUniformF}
f \left( b_1, b_0, \ldots, b_0 \right) = q_1 f \left( b_0, \ldots, b_0 \right)
\end{align}
%
for some arbitrary
constant $q_1$. \ednote{It is unclear why this can be/must be rational}
Then we can substitute \ednote{deleted `as from'} \eqref{RefIteration0} inside \eqref{RefIteration1}, obtaining a rational number $k_1$ not depending on $b_0, b_1$ such that
\begin{align}
P \left( \Set{
b_1, \underbrace{b_0, \ldots, b_0}_{N-2}
} \right) = k_1 f \left( b_0, \ldots, b_0 \right).
\end{align}
Proceding the same way, we can build a rational constant $k_2$ such that
\begin{align}
P \left( \Set{
b_1, b_2, \underbrace{b_0, \ldots, b_0}_{N-3}
} \right) = k_2 f \left( b_0, \ldots, b_0 \right)
\end{align}
for arbitrary $b_1, b_2$.

So that by iterating this mechanism we can construct a relation binding the generic 
$P \left( \Set{b_1, \ldots, b_{N-2}, b_0 }\right)$ to 
$f \left( b_0, \ldots, b_0 \right)$:
\begin{align}
\label{RefEqForContradiction}
P \left( \Set{b_1, \ldots, b_{N-2}, b_0 }\right) = k_{N-2} 
f \left( b_0, \ldots, b_0 \right).
\end{align}
Moreover, at each step of this iteration, the requirement \eqref{RefUniformF} gives indications about how $q_1, q_2, \ldots$, $X$ and $f$ must be related if we want the proof mechanism to work.
It is important to note that, in doing so, such mutual relations should be weakened as much as possible, with the only rationale that they should grant that the proof mechanism just outlined actually works.
For example, we imposed one equality of the kind \eqref{RefUniformF} at each inductive step, but these equalities actually need to hold only for the bid vectors inside some minimal $X$; otherwise, we would restrict ourselves to quite trivial instances of $f$.
In this section, we wanted to give a general idea of the proof, and we did not explicitly state the exact mutual relations between $b_0$, $X$ and $f$.
Indeed, such relations are not immediate, at first: they actually became clearer when formalizing the proof itself; a process that we feel would have been harder to manage with a standard paper-based proof.
These relations will be given in full detail \ednote{replaced `details' by `detail'} in Section \ref{RefSectLemmaInductive}, in Definition \ref{RefDefAdequate}.

The iteration explained above implies that we assign some value to the numbers $q_1, q_2, \ldots$.
We deemed them arbitrary because the proof mechanism works whichever values we assign them.
For simplicity, however, we restricted our work to the case 
\begin{align}
\label{RefAssumptSimplicity}
1=q_1=q_2=\ldots,
\end{align}
which will be general enough for any application to auction theory.

The idea is now to take equation \eqref{RefEqForContradiction}, which was obtained using equation \eqref{RefAlgebraicProblSymmetric}, and to derive a contradiction between it and \eqref{RefAlgebraicProblSymmetric} itself.
Hence, the formalization can be seen as split in two stages: there is Lemma \ref{RefLemmaInductive}, presented in Section \ref{RefSectLemmaInductive}, which formalizes equation \eqref{RefEqForContradiction} and takes care of spelling out all the exact requirements in order to derive it exploiting the idea we informally presented above.
Then there are other auxiliary definitions and lemmas, presented in Section \ref{RefSectTheorem}, which employ Lemma \ref{RefLemmaInductive} to obtain the wanted contradiction (given in the thesis of Theorem \ref{RefThmNbFormal}).

\section{From the idea to the formal statement}
\label{RefSectLemmaInductive}
Given a multiset $m$, an $m$-restriction of $\v b$ is any $\v b' \subseteq \v b$ such that $\RRange {\v b'} = m$.

An $m$-completion of $\v b$ to $b_0$ is a $\v b'$ writable as the disjoint union of an $m$-restriction of $\v b$ with a function constantly valued $b_0$, and such that $\dom \v b = \dom \v b'$.
In other words, an $m$-completion of $\v b$ to $b_0$ is obtained from $\v b$ by changing its value to $b_0$ outside some $m$-restriction of it.

A full family of $b_0$-completions of $\v b$ is a set containing one $m$-completion of $\v b$ to $b_0$ for each possible $m \msub \RRange{\v b}$.

\begin{lemma}[\texttt{lll57}]
\label{RefLemmaInductive}
Consider a full family $Y$ of $b_0$-completions of $\v b$, and set 
$X := \left( \left\{ i_1, i_2 \right\} \times \left\{ b_0 \right\} \right) \adjoin Y$
for some $i_1 \neq i_2$ outside $\dom \v b$.
Assume that, $\forall \v b' \in X$:
\begin{align}
\label{RefHyp1}
f \left( \v b' \right) &= f \left( \left( \left\{ i_1, i_2 \right\} \cup \dom \v b \right) \times \left\{ b_0 \right\}
 \right)
\\[2mm]
\label{RefHyp2}
f \left( \v b' \right) &= \sum_{i \in \dom \v b'} P \left( \v b' \out i \right).
\end{align}
Then 
\begin{align*}
P \left( \RRange {\v b} \mcup \Set{b_0} \right) = 
\frac{1}{2+\card{\dom \v b}} 
f \left( 
\left( 
\left\{ i_1, i_2 \right\} \cup \dom \v b 
 \right)
\times \left\{ b_0 \right\} \right).
\end{align*}
\end{lemma}

For later discussion, it will be useful to express requirements in Lemma \ref{RefLemmaInductive} up to equality \eqref{RefHyp1} in a predicate form:
\begin{definition}
\label{RefDefAdequate}
[\texttt{pred2, pred3}]
\sloppy{}
The set $X$ is \emph{\pred{}} for the quintuple $\left( \v b, b_0, f, i_1, i_2 \right)$ if
\begin{itemize}
\item
$ X = \left\{ i_1, i_2 \right\} \adjoin{} Y$ for some $Y$ being a full family of $b_0$-completions of $\v b;$
\item
$\forall \v b' \in X \ \ f \left( \v b' \right) = f \left( \left( \left\{ i_1, i_2 \right\} \cup \dom \v b \right) \times \left\{ b_0 \right\}
 \right).$
\end{itemize}
\end{definition}
\fussy
This allows us to restate Lemma \ref{RefLemmaInductive} as 

\begin{lemma}
[\texttt{lll57}]
\label{RefLemmaInductive2}
Assume that $X$ is \pred{} for the quintuple $\left( \v b, b_0, f, i_1, i_2 \right)$, and that 
\begin{align*}
\forall \v b' \in X && \sum_{i \in \dom \v b'} P \left( \v b' \out i \right) = f \left( \v b' \right).
\end{align*}
Then
\begin{align*}
P\left( \RRange{\v b} \mcup \Set{b_0} \right) 
=
\frac {1} {2 + \card{\dom \v b}} f \left( 
\left( \left\{ i_1, i_2 \right\} \cup \dom \v b  \right)
\times \left\{ b_0 \right\}
\right).
\end{align*}
\end{lemma}

\section{Example application to the Vickrey auction}
\label{RefSectVickrey}
Let us consider the specific case of $f = - f_2$: we recall that $f_2 \left( \v b \right)$ is the maximum of the bids $\v b$, once the bid of the winner has been removed. 
In this case, choosing any
$b_0 \geq \max (\ran {\v b})$ satisfies hypothesis \eqref{RefHyp1} of Lemma \ref{RefLemmaInductive}, permitting 
\begin{align}
\label{RefCorollarySecondPrice}
P \left( \RRange {\v b} \mcup \Set{b_0} \right) =
\frac{- b_0}{2+\card{\dom \v b}}.
\end{align}
Let us apply this to two particular bid vectors:
\begin{align*}
\underline {\v b} := \left( 1, 2, \ldots, n, n+1, n+3 \right) 
&& 
\overline{\v b} := \left( 1, 2, \ldots, n, n+2, n+3 \right).
\end{align*}
We get
\begin{multline}
\label{RefInequality}
f_2 \left( \underline {\v b} \right) + \sum_{i} P \left( \RRange{\underline {\v b} - i} \right) 
\overset{\text{\tiny{\eqref{RefCorollarySecondPrice}}}}
{=} 
\left( n+1 \right) 
- 
\left[ \frac{\left( n+3 \right)}{n+2} \left( n+1 \right) \right]
- \frac{n+1}{n+2}
\\[2mm]
\neq
n+2 - 
\left[ \frac{\left( n+3 \right)}{n+2} \left( n+1 \right) \right]
- \frac{n+2}{n+2}
\overset{\text{\tiny{\eqref{RefCorollarySecondPrice}}}}
{=}
f_2 \left( \overline{\v b} \right) + \sum_{i} P \left( \RRange{\overline{\v b} - i} \right).
\end{multline}
Thus, 
we have falsified \eqref{RefBalanceDefParticular} as an application of Lemma \ref{RefLemmaInductive}.
To do that, we had to apply Lemma \ref{RefLemmaInductive} $n+2$ times for the\ednote{added the} first equality of the chain above, and further $n+2$ times for its last equality.
This corresponds to having imposed \eqref{RefAlgebraicProblSymmetric} on the sets
\begin{align}
\label{RefFamily1}
\left\{ \left\{ i, j \right\} \times \left\{ n+3 \right\} \adjoin \completion{\underline {\v b} \out i \out j}{n+3}
\right\}_{
\substack{\underline {\v b} \left( j \right)=n+3\\i \in \dom {\underline {\v b}} - \left\{ j \right\}}}
\\[1mm]
\label{RefX1}
\left( {\underline {\v b}}^{-1}\left\{ n+1, n+3 \right\} \right) 
\times \left\{ n+1 \right\} \adjoin
\completion{\underline {\v b} \out 
{\underline{\v b}}^{-1} \left\{ n+1, n+3 \right\}}{n+1}
\end{align}
and on the sets
\begin{align}
\label{RefFamily2}
\left\{ \left\{ i, j \right\} \times \left\{ n+3 \right\} \adjoin \completion{\overline {\v b} \out i \out j}{n+3}
\right\}_{
\substack{\overline {\v b} \left( j \right)=n+3\\i \in \dom {\overline{\v b}} - \left\{ j \right\}}}
\\[2mm]
\label{RefX2}
\left( {\overline {\v b}}^{-1}\left\{ n+2, n+3 \right\}  \right)
\times \left\{ n+2 \right\} \adjoin
\completion{\overline {\v b} \out 
{\overline {\v b}}^{-1} \left\{ n+2, n+3 \right\}}{n+2}
\end{align}
\ednote{deleted comma}respectively.
Above, we have indicated with $\completion{\v b}{b}$ a fixed, arbitrary full family of $b$-completions of $\v b$.
Hence, the union of the family of sets in \eqref{RefFamily1}, the union of that in \eqref{RefFamily2}, the sets in \eqref{RefX1}, \eqref{RefX2}, together with $ \left\{ \underline{\v b}, \overline{\v b} \right\}$, form the wanted set $X$: we have a contradiction when imposing \eqref{RefAlgebraicProblSymmetric} on it.

\section{Application to the general case}
\label{RefSectTheorem}
Formally, as a first step of what we did in Section \ref{RefSectVickrey}, we apply Lemma \ref{RefLemmaInductive} to each possible $\v b \out i$ appearing in \eqref{RefAlgebraicProblSymmetric}, obtaining an equality for the sum featured there.
The following result does exactly that and is an immediate corollary of Lemma~\ref{RefLemmaInductive2}:\ednote{deleted comma}

\begin{corollary}
[\texttt{lll68}]
\label{RefCorollary}
Assume that $ \forall i \in \dom \v b $ there are $j_i$ and $X_i$ such that 
\begin{align*}
j_i \in \dom \v b \sdiff \left\{ i \right\}
\\
X_i \text{ is \pred{} for} 
\left( \v b \out \left\{ i, j_i \right\}, \v b \left( j_i \right), f, i, j_i \right).  
\end{align*}
Also assume that 
\begin{align*}
\forall \v b' \in \bigcup X_i && 
\sum_{i \in \dom \v b'} P \left( \RRange{\v b' \out i} \right) = f \left( \v b' \right).
\end{align*}
Then
\begin{align}
\label{RefCorollaryThesis}
\sum_{i \in \dom \v b} P \left( \Set{\v b \out i} \right) = 
\frac{1}{\card{\dom \v b}} 
\sum_{i \in \dom \v b} f \left( \dom \v b \times \left\{ \v b \left( j_i \right) \right\}\right).
\end{align}
\end{corollary}

What we informally did in Section \ref{RefSectVickrey} was to find $\underline {\v b}$, $\overline {\v b}$ to each of which\ednote{deleted `the'} corollary~\ref{RefCorollary} is applicable, but such that the maps
$\underline \eta: i \mapsto f \left( \dom {\underline{\v b}} \times \left\{ \underline {\v b} \left( \underline j_i \right) \right\} \right)$ and 
$\overline \eta: i \mapsto f \left( \dom {\overline{\v b}} \times \left\{ \overline {\v b} \left( \overline j_i \right) \right\} \right)$ enjoy the following properties:
\begin{enumerate}
\item
\label{RefFact1}
each assumes exactly two values, call them $\underline{v_1} \neq \underline{v_2}$ and $\overline{v_1} \neq \overline{v_2}$, respectively;
\item
\label{RefFact2}
of these four values, exactly two are equal, let us say $\underline{v_1}= \overline{v_1}$, while 
$\underline{v_2} \neq \overline{v_2}$;
\item
\label{RefFact3}
the sets $ {\underline \eta}^{-1} \left\{ \underline {v_1} \right\}
$ and $  {\overline \eta}^{-1} \left\{ \underline {v_1} \right\}$ coincide: that is, the points on which $\underline \eta$ and $\overline \eta$ yield the same value are exactly the same.
\end{enumerate}
These facts cause the occurrence of the two identical terms which can be cancelled\ednote{cancelled instead of canceled} in expression \eqref{RefInequality}: they are put in square brackets there for clarity. 
This cancellation is fundamental, because it immediately allows to establish the inequality appearing there.
Finding such $\underline{\v b}$ and $\overline{\v b}$ is particularly easy in the case of $f = f_2$, but the same mechanism works for a generic $f$, leading to a similar cancellation between the two sums
\begin{align*}
\sum_{i \in \dom \underline{\v b}} f \left( \dom {\underline {\v b}} \times \left\{ \underline{ \v b} \left( \underline j_i \right) \right\}\right)
\end{align*}
and
\begin{align*}
\sum_{i \in \dom \overline{\v b}} f \left( \dom {\overline {\v b}} \times \left\{ \overline{ \v b} \left( \overline j_i \right) \right\}\right);
\end{align*}
each of those sums is yielded by a separate application of corollary \ref{RefCorollary}, in the right hand side of equation \eqref{RefCorollaryThesis}.

To formalize the requirements \eqref{RefFact1}, \eqref{RefFact2}, \eqref{RefFact3} appearing in the list above, we introduce the following definition:

\begin{definition}
[\texttt{counterexample}]
\label{RefDefCounterexample}
The triple $ \left( \underline {\v b}, \overline{\v b}, h \right)$ is a \emph{\cex{}} for the map $f$ if
$ \dom {\ul {\v b}} = \dom {\ol {\v b}}$ and there is a map $ g $ such that
\begin{align*}
h \colon \left( \dom {\ul {\v b}}  \right) \to \left\{ f, g \right\}
\\
\left\{ 0 \right\} \subset
\left\{ 
f \left( \ol{\v b} \right) - f \left( \ul{\v b} \right),
g \left( \ol{\v b} \right) - g \left( \ul{\v b} \right)
\right\}. 
\end{align*}
\end{definition}
This definition is devised exactly to formulate the following lemma, which is an easy arithmetical statement:
\begin{lemma}
[\texttt{lll69}]
\label{RefLemmaInequality}
Assume that $ \left( \ul{\v b}, \ol{\v b}, h \right) $ is a \cex{} for $f$, and that $ 2 \leq \card{\dom \ul{\v b}} < + \infty$.
Then
\begin{align*}
f \left( \ul{\v b} \right) - \frac{1}{\card{ \dom \ul{\v b} }} \sum \left( h \left( i \right) \right) \left( \ul {\v b} \right)
\neq
f \left( \ol{\v b} \right) - \frac{1}{\card{ \dom \ol{\v b} }} \sum \left( h \left( i \right) \right) \left( \ol {\v b} \right).
\end{align*}
\end{lemma}
In turn, the lemma above can finally be combined with corollary \ref{RefCorollary} into the main theorem:

\begin{theorem}
[\texttt{tt01}]
\label{RefThmNbFormal}
Assume that $\left( \underline {\v b}, \overline{\v b}, h \right)$ is a \cex{} for $f$, with $ 2 \leq \card {\dom \ul{\v b}} < + \infty $.
Moreover, assume that \\ 
$ \forall i \in \dom \ul{\v b}$ there are $\ul j_i$, $\ul X_i$, $\ol j_i$, $\ol X_i$ satisfying
\begin{align*}
\ul j_i \in \dom \ul{\v b} \sdiff \left\{ i \right\}
\\
\ul X_i \text{ is \pred{} for} 
\left( \ul{\v b} \out \left\{ i, \ul j_i \right\}, \ul{\v b} \left( \ul j_i \right), f, i, \ul j_i \right)
\\
f\left( \dom \ul{\v b} \times \left\{ 
\ul{\v b} \left( \ul j_i \right)
\right\}  \right)
=\left( h \left( i  \right) \right) \left( \ul{\v b} \right)
\\
\ol j_i \in \dom \ol{\v b} \sdiff \left\{ i \right\}
\\ 
\ol X_i \text{ is \pred{} for} 
\left( \ol{\v b} \out \left\{ i, \ol j_i \right\}, \ol{\v b} \left( \ol j_i \right), f, i, \ol j_i \right)
\\
f\left( \dom \ol{\v b} \times \left\{ 
\ol{\v b} \left( \ol j_i \right)
\right\}  \right)
=\left( h \left( i  \right) \right) \left( \ol{\v b} \right)
. 
\end{align*}
Finally, suppose that
\begin{align*}
\forall \v b' \in \bigcup_{i \in \dom {\ul{\v b}}} \ul X_i \cup \ol X_i 
&&
\sum_{k \in \dom \v b'} P \left( \RRange{\v b' \out k} \right) = f \left( \v b' \right).
\end{align*}
Then 
\begin{align*}
f \left( \ul{\v b} \right) - \sum_{i \in \dom{\ul{\v b}}} P\left( \RRange{\ul{\v b} \out i} \right)
\neq
f \left( \ol{\v b} \right) - \sum_{i \in \dom{\ol{\v b}}} P\left( \RRange{\ol{\v b} \out i} \right).
\end{align*}
\end{theorem}

\section{Conclusions}\label{RefConclusions}

We developed a result characterizing imbalanced auction mechanisms.
Both the theorem and the proof are new to the best of our knowledge, and we informally illustrated the idea behind the proof.

On the formal side, the proof has been implemented in the \I{}/HOL theorem prover, which is especially important in this case, because the confidence added by the formalization is a significant soundness validation for any new result.

Given a rather general class of auction mechanisms, our theorem provides explicit conditions implying the imbalance condition, and in this sense it can also be regarded as a result in reverse game theory.
Moreover, our theorem also explicitly constructs a finite set on which the imbalance condition holds: this can be exploited in concrete implementations to computationally check the imbalance condition only over that known, finite set.

The fact that the proof and the result are new also leaves open many avenues for possible generalizations and improvements.
For example, assumption \eqref{RefAssumptSimplicity} \ednote{delete reference to pages. They may go wrong `on page \pageref{RefAssumptSimplicity}'} was taken for the sake of simplicity, but less immediate assumptions are also possible.
Similarly, Definition \ref{RefDefCounterexample} is merely the simplest one granting that Lemma \ref{RefLemmaInequality} holds: there are plenty of ways to achieve the same result, which can lead to different final requirements on $f$ appearing in the statement of Theorem~\ref{RefThmNbFormal}.
Currently, \fo{} is following these tracks to investigate further developments of possible interest to its application domain, auction theory.

\printbibliography
\end{document}